\documentstyle[graphics,onecolumn]{mn}
\newcommand{\mincir}{\raise
-2.truept\hbox{\rlap{\hbox{$\sim$}}\raise5.truept 
\hbox{$<$}\ }}
\newcommand{\magcir}{\raise
-2.truept\hbox{\rlap{\hbox{$\sim$}}\raise5.truept
\hbox{$>$}\ }}
\newcommand{\minmag}{\raise-2.truept\hbox{\rlap{\hbox{$<$}}\raise
6.truept\hbox
{$>$}\ }}

\newcommand{\be}{\begin{equation}}
\newcommand{\ee}{\end{equation}}
\newcommand{\ba}{\begin{eqnarray}}
\newcommand{\ea}{\end{eqnarray}}
\newcommand{\brr}{\begin{array}}
 
\newcommand{\err}{\end{array}}
\newcommand{\bc}{\begin{center}}
\newcommand{\ec}{\end{center}}

\newcommand{\et}{{\it et al.~}}

\title[Detecting X-ray filaments in
the low redshift Universe] 
{Detecting X-ray filaments in
the low redshift Universe with {\it XEUS} and 
{\it Constellation-X}}
\author[M. Viel, E. Branchini, R. Cen, S. Matarrese, P. Mazzotta \&
J.P. Ostriker] 
{M. Viel $^{1, 2}$, E. Branchini $^{3}$, R. Cen $^4$, S. Matarrese
$^{1, 5}$, P. Mazzotta $^6$ \& \cr J.P. Ostriker $^{2}$
\\ 
$^1$ Dipartimento di Fisica `Galileo Galilei',
via Marzolo 8, I-35131 Padova, Italy \\
$^2$ Institute of Astronomy, Madingley Road, Cambridge CB3 0HA, UK\\
$^3$ Dipartimento di Fisica, Universit\`a di Roma TRE, Via della Vasca
Navale 84, 00146, Roma, Italy\\
$^4$ Princeton University Observatory, Princeton University, Princeton NJ 08544 \\
$^5$ INFN, Sezione di Padova, via Marzolo 8, I-35131 Padova, Italy \\
$^6$ Department of Physics, University of Durham, South Road,
Durham, DH1 3LE, UK\\
Email: viel@pd.infn.it, branchini@fis.uniroma3.it, cen@astro.princeton.edu, matarrese@pd.infn.it, mazzotta@durham.ac.uk,\\ \hspace{.8cm} jpo@ast.cam.ac.uk}

\date{\bf accepted by MNRAS}

\begin{document}

\maketitle

\begin{abstract}
We propose a possible way to detect baryons at low redshifts from the
analysis of X-ray absorption spectra of bright AGN pairs. A simple
semi-analytical model to simulate the spectra is presented. We model
the diffuse warm-hot intergalactic medium (WHIM) component, responsible
for the X-ray absorption, using inputs from high-resolution
hydro-dynamical simulations and analytical prescriptions.  We show that
the number of OVII absorbers per unit redshift with column density
larger than $10^{13.5}$ cm$^{-2}$ - corresponding to an equivalent
width of $\sim$ 1 km/s - which will be possibly detectable by {\it
XEUS}, is $\magcir 30$ per unit redshift. {\it Constellation-X} will
detect $\sim 6$ OVII absorptions per unit redshift with an equivalent
width of 10 km/s. Our results show that, in a $\Lambda$CDM Universe,
the characteristic size of these absorbers at $z\sim 0.1$ is $\sim 1$
$h^{-1}$ Mpc.  The filamentary structure of WHIM can be probed by
finding coincident absorption lines in the spectra of background AGN
pairs. We estimate that at least 20 AGN pairs at separation $\mincir
20$ arcmin are needed to detect this filamentary structure at a
3$\sigma$ level.  Assuming observations of distant sources using {\it
XEUS} for exposure times of 500 ksec, we find that the minimum source
flux to probe the filamentary structure is $\sim 2\times 10^{-12}$ erg
cm$^{-2}$ s$^{-1}$, in the 0.1-2.4 keV energy band. Thus, most pairs of
these extragalactic X-ray bright sources have already been identified
in the {\it ROSAT} All-Sky Survey. Re-observation of these objects by
future missions could be a powerful way to search for baryons in the
low redshift Universe.
\end{abstract}

\begin{keywords}
Cosmology: theory -- intergalactic medium -- large-scale structure of
universe -- quasars: absorption lines
  
\end{keywords}

\section{Introduction}
\label{intro}

The census of baryons in the low redshift Universe shows that a large
fraction of them has not been detected yet. While almost 80\% of the
baryons reside in the Ly$\alpha$ forest at $z\sim 2$ (Rauch 1998),
estimates at lower redshift imply that the observed baryon density is
significantly lower (Fukugita {\it et al.} 1998; Nicastro {\it et al.}
2002b) and does not match the nucleosynthesis constraints (Burles \&
Tytler 1998).  Hydro-dynamical simulations suggest that a large
fraction of baryons at $z \sim 0$ could be in a warm-hot intergalactic
medium (WHIM), with a temperature between $10^5$ and $10^7$ K and with
moderate overdensities $\delta\sim 10-100$ (Ostriker \& Cen 1996; Cen
\& Ostriker 1999; Dav\'e {\it et al.} 2001; Cen {\it et al.} 2001). 
Detection of this component is particularly difficult and requires
very sensitive UV and X-ray satellites.

Recently, various attempts to model analytically the WHIM component
have been made.  Perna \& Loeb (1998) used the Press \& Schechter
theory (Press \& Schechter 1974) and an isothermal model for the gas
component to address the detectability in absorption of the hot and
highly ionized intergalactic medium (IGM), which resides in the
outskirts of clusters and groups of galaxies. Valageas, Schaeffer \&
Silk (2002) showed that the temperature-density relation found in
hydro-dynamical simulations can be reproduced using straightforward
physical arguments. In this simple picture the IGM consists of two
phases: the cool photo-ionized component which gives rise to the
Ly$\alpha$ forest and the WHIM, dominated by shock heating.

Hydro-dynamical simulations of structure formation in the context of
cold dark matter models have been the major contribution to the
theoretical understanding of the WHIM.  Hellsten {\it et al.}  (1998),
referred to the network of filaments and sheet-like structures,
possibly seen in the spectra of background bright sources in the X-ray
band, as the `X-ray forest', in analogy with the Ly$\alpha$
forest. Their main result is that OVII and OVIII absorptions by the
intervening IGM can be seen by future X-ray missions.  Cen {\it et al.}
(2001) used large box-size hydro-dynamical simulations to detect OVI
absorption by the WHIM in the UV band and find results in reasonably
good agreement with observations.  Fang {\it et al.}  (2002a) performed
hydro-dynamical simulations to study the metal distribution in the IGM
and, assuming collisionally ionized gas, compared their results with
semi-analytical models based on the Press \& Schechter
formalism. Kravtsov {\it et al.} 2002 addressed WHIM detectability
using hydro-dynamical simulations to model the gas distribution in the
Local Supercluster region and imposing constraints from the MARK III
catalogue of galaxy peculiar velocity.

More recently, Chen {\it et al.} (2002) have made an extensive study of
the detectability of OVII and OVIII absorbers using the $z=0$ output of
a hydro-dynamical simulation of a $\Lambda$CDM Universe, and assuming a
uniform metallicity of 0.1 solar for the IGM. Their results confirm
that the detectability of these absorbers is challenging.  It has been
shown that the number of intervening OVI absorbers per unit redshift is
quite high at low redshift (Tripp \et 2000).  Thus, a very promising
technique to detect OVII and OVIII systems is to search for
corresponding UV OVI absorptions in the spectra: this allows to lower
the S/N ratio needed for OVII and/or OVIII detectability. Nicastro {\it
et al.}  (2002a) presented the X-ray 5$\sigma$ level detection by {\it
Chandra} of a resonant absorption from WHIM along the line of sight
towards the blazar PKS 2155-304. Moreover, Fang {\it et al.}  (2002b)
reported a detection of an OVIII absorption line along the sight-line
towards PKS 2155-304 with {\it Chandra}. They constrained the gas that
gives rise to this line to have an overdensity in the range 30-350 and
a temperature of 4-5 $\times 10^6$ K, in agreement with
hydro-simulations. Also emission by the WHIM could be important and
several models have been proposed (e.g. Scharf {\it et al.}  2000; 
Phillips {\it et al.} 2001; Zappacosta {\it et al.} 2002). Here, we will
focus on absorption: the bulk of the missing baryons should reside in a
medium which is too rarefied to be detectable through its X-ray
emission or Sunyaev-Zel'dovich effect (Ostriker \& Cen 1996).

These very recent observations are very encouraging and show that we have
just started to detect the first `trees' of the X-ray forest. Future
X-ray missions such as {\it XEUS} and {\it Constellation-X} will be 
much more accurate a probe of the current cosmological models and will
allow a detailed study of the WHIM properties.

In this paper we will present a semi-analytical technique to simulate
X-ray absorption spectra. There are two advantages in using a
semi-analytical technique instead of the more accurate hydro-dynamical
simulations. First, semi-analytical models are not computationally
expensive and are not limited by resolution and box-size
effects. Second, semi-analytical techniques can be very efficient when
coupled with detailed hydro-dynamical simulations when the latter can
be used to calibrate the former and allow one to explore very quickly
the parameter space. For example, it will be straightforward to study
the effects of the metallicity of the IGM, different UV and X-ray
backgrounds, scatter in the temperature-density and in the
metallicity-density relation, peculiar velocities etc.  For the
estimates we want to make here, semi-analytical techniques, which use
the inputs of hydro-dynamical simulations, will show to be
successful in reproducing potentially observable quantities such as
the column densities of OVII and OVIII absorbers.

In particular, we will focus on the simulations of absorption spectra
of AGN pairs and we will try to determine the characteristic size of
the absorbers by identifying absorption features in the simulated
spectra.  The estimate of the extent of the absorbers, using the
information contained in the transverse direction, could be a very
powerful probe of the filamentary network of WHIM.  An extensive
analysis of the correlations of Ly$\alpha$ clouds at $z>2$ using
hydro-dynamical simulations has been made in Miralda-Escud\'e \et
(1996) and Cen \& Simcoe (1997). McGill (1990), Crotts \et (1994),
Bechtold \et (1994), Charlton {\it et al.} (1997) performed similar
analysis for observed Ly$\alpha$ clouds and showed that the coherence
length in the transverse direction of the clouds could be significantly
larger than the size of a single cloud. This argument has been
clarified recently using simple physical arguments (Schaye 2001) and
hydro-dynamical simulations (e.g. Theuns {\it et al.}  1998), which
indeed show that the clouds are embedded in a network of filaments.
The idea is very simple: while a single line-of-sight (LOS) allows to
probe only the size of a single absorber, pairs of LOS are sensitive to
the random orientation of the filaments and diffuse blobs and can
better sample the WHIM.  Another advantage of using the information
contained in a sample of pairs is that this method minimizes the
contamination of the WHIM signal we have to detect with the intrinsic
lines of the background source. This is due to the fact that the
positions of the intrinsic absorption or emission features of the
source along the LOS are random. Thus, their effect can be
statistically modelled and removed as it has been done recently to
recover the linear dark-matter power spectrum from simulated Ly$\alpha$
absorptions (Viel {\it et al.} 2002a).

The outline of the paper is as follows. In Section 2 we describe the
ingredients needed to simulate the X-ray AGN spectra focussing both on
the inputs from hydro-dynamical simulations (subsection
\ref{hydro}) and on the analytical modelling (subsection
\ref{sam}). In Section 3 we test how well the semi-analytical model
matches the properties of simulated lines extracted directly from
hydro-simulations.  This comparison is done in terms of the predicted
number counts of absorbers per unit redshift range. In Section 4 we
discuss a possible way to recover the characteristic size of these
absorbers by using the so called `hits-and-misses' statistics and
discuss detectability of these absorption features by satellites
of the next generation such as {\it XEUS} and {\it
Constellation-X}. Section 5 contains our main conclusions.

\section{Semi-analytical model of the X-Ray forest}
This Section presents the semi-analytical method used to simulate
X-ray spectra of AGN.  
In Subsection \ref{hydro} we will present the hydro-dynamical
simulation used, focussing on the basic ingredients which will
constitute the inputs for the semi-analytical model: {\it i)} the
probability distribution function (pdf) of the gas;
{\it ii)} the gas metallicity - density relation; {\it iii)} the
gas temperature - density relation; {\it iv)} the X-ray
background.  In Subsection \ref{sam} we will describe in detail the
technique used, with the relative mathematical implementation
(see also Appendix A1), to produce the simulated spectra of AGN pairs.

\subsection{Model parameters from hydro-dynamical simulations}
\label{hydro}
The hydro-dynamical simulation used here is a $\Lambda$CDM model with
the following parameters: $H_0=100h$ km s$^{-1}$Mpc$^{-1}$ with $h=0.67$,
$\Omega_{0m}=0.30$, $\Omega_{0b}=0.035$, $\Lambda=0.70$, $\sigma_8=0.90$,
and the spectral index of the primordial power spectrum $n=1$. The
box-size of the simulation is $25$ comoving $h^{-1}$ Mpc on a uniform
mesh with $768^3$ cells. The comoving cell size is 32.6 $h^{-1}$ kpc
and the mass of each dark matter particle is $\sim 2 \times 10^7$
M$_{\odot}$ (further details can be found in Cen {\it et al.} 2001).
The simulation includes galaxy and star formation, energy feedback
from supernova explosions, ionization radiation from massive stars and
metal recycling due to SNe/galactic winds. Metals are ejected into the
local gas cells where stellar particles are located using a yield
$Y=0.02$ and are followed as a separate variable adopting the standard
solar composition (some slices through the simulation box at various
redshift outputs can be found at
http://astro.Princeton.EDU/$\sim$cen/PROJECTS/p2/p2.html).

\begin{figure*}
\center
\resizebox{.5\textwidth}{!}{\includegraphics{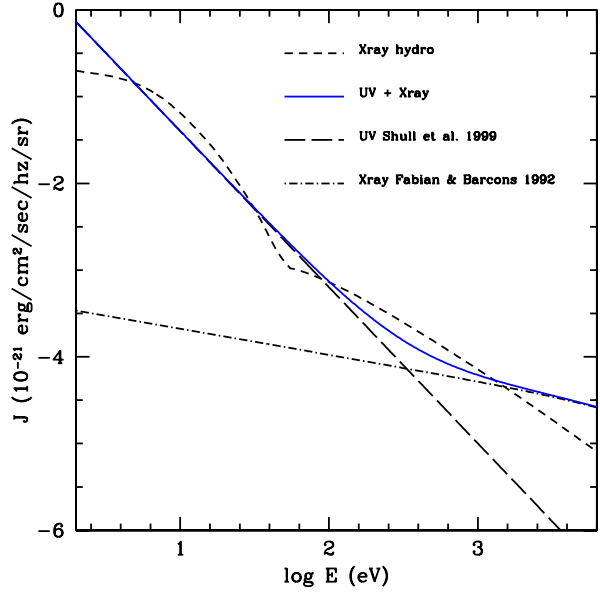}} 
\caption{The X-ray and UV background. The short-dashed line shows the
background extracted from the $z=0$ output of the hydro-dynamical
simulations and used to compute our spectra. The estimates of Shull
{\it et al.}  (1999) for the UV background are represented by the
long-dashed line. The results of Fabian \& Barcons (1992) for the X-ray
background are also reported (dot-dashed line). The sum of the latter
two is represented by the continuous line.}
\label{back}
\end{figure*}

To properly simulate X-ray absorptions we have to know the actual X-ray
and UV background of the simulation at $z=0$, which is computed
self-consistently given the sinks and the sources in the simulation
box.  We have checked that there are very small differences between the
hydro-simulation background and the estimate from Shull {\it et al.}
(1999): $I_{UV} = I_{UV}^0 (E/13.6\rm{eV})^{-1.8}$, with
$I_{UV}^{0}=2.3 \times 10^{-23}$ erg cm$^{-2}$ Hz$^{-1}$ sr$^{-1}$
s$^{-1}$ for the UV background (which includes a contribution from AGNs
and starburst galaxies), and
$I_{X}=I_{X}^{0}(E/E_{X})^{-1.29}\exp{(-E/E_{X}) }$, with
$I_{X}^{0}=1.75 \times 10^{-26}$ erg cm$^{-2}$ Hz$^{-1}$ sr$^{-1}$
s$^{-1}$ and $E_{X}=40$ keV (Boldt 1987; Fabian \& Barcons 1992;
Hellsten {\it et al.} 1998; Chen {\it et al.} 2002), for the X-ray
background.  The difference between the amplitude of the UV and X-ray
background of the hydro-simulations and the estimates above is less
than a factor of 2, in the range 0.003 \mincir E (keV) \mincir 4.  In
Figure \ref{back} we plot the X-ray and UV background of the
hydro-simulations at $z=0$ (dashed line), the estimates of Shull {\it
et al.}  (1999) (long dashed line) and Fabian \& Barcons 1992
(dot-dashed line), and the sum of these two (continuous line).

Among the two backgrounds, the X-ray one plays a major role in
determining the statistics of the simulated spectra.  We will assume
that the amplitude of the ionizing backgrounds scales like $(1+z)^3$,
in the redshift range $0<z<1$. This assumption will not influence
significantly our results: given the fact that we will generate
mock-spectra at $z<0.3$, this will determine a scaling of the
amplitude by a factor of less than 2.

\begin{figure*}
\center
\resizebox{.5\textwidth}{!}{\includegraphics{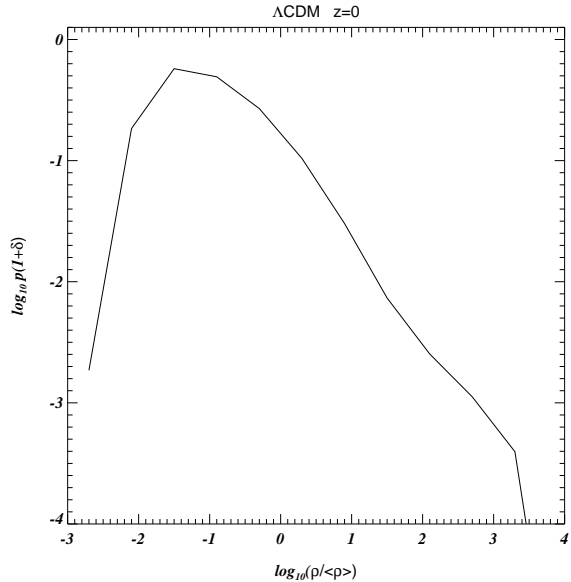}} 
\caption{Probability distribution function of the gas overdensity for
the $z=0$ output of the $\Lambda$CDM model.}
\label{fig1}
\end{figure*}
\begin{figure*}
\center
\resizebox{.5\textwidth}{!}{\includegraphics{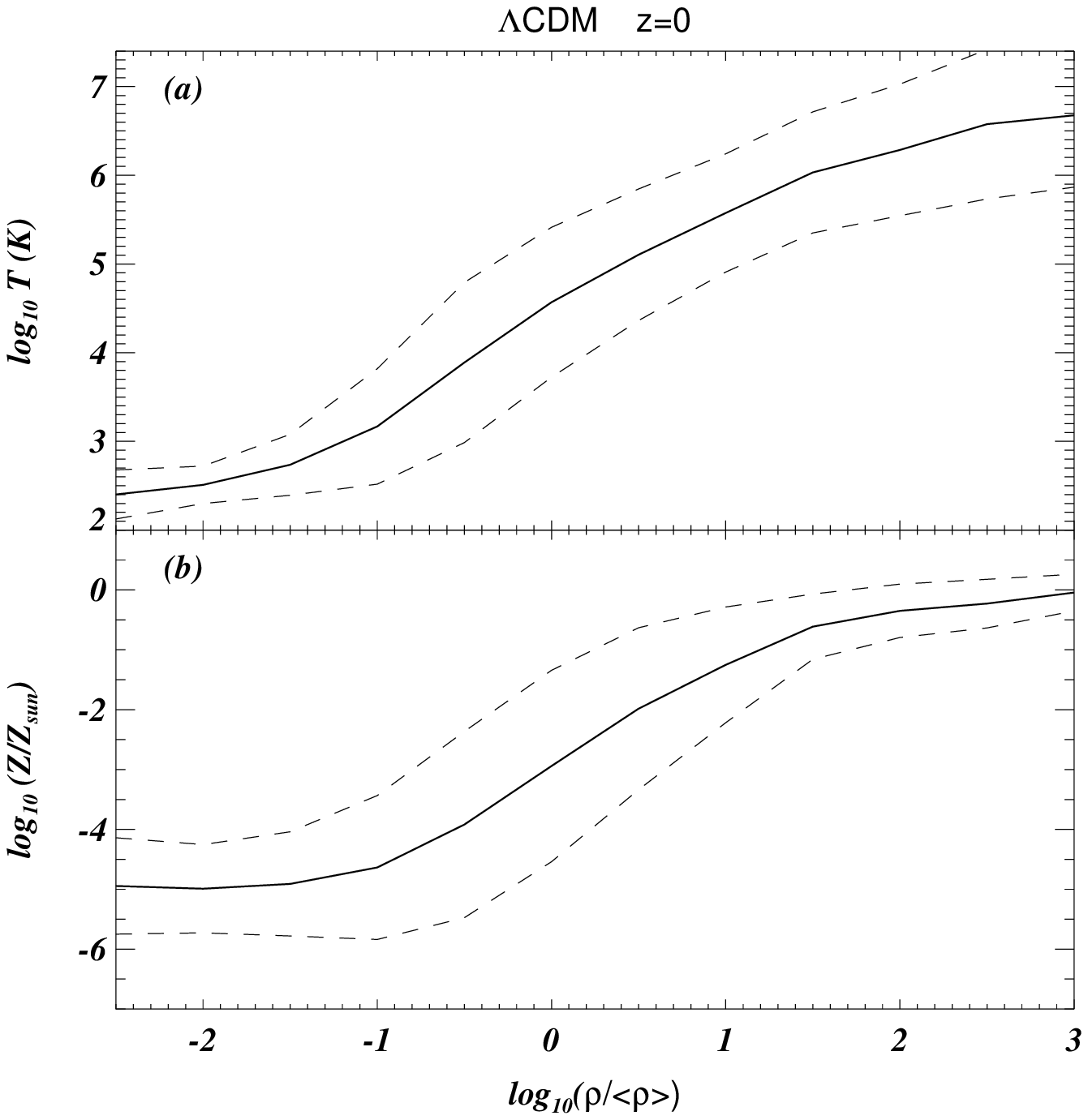}} 
\caption{Panel $(a)$ shows the temperature density relation for the gas 
component, while panel $(b)$ shows the metallicity as a function of
density. The continuous lines represent the mean value in each bin, the dashed
lines represent the scatter (1$\sigma$ values). The quantities shown are 
volume weighted.}
\label{fig2}
\end{figure*}
Along many randomly chosen sight-lines parallel to one of the axes of
the simulation box, we compute the gas density and peculiar velocity,
the gas temperature, and the metal density. This allows us to obtain 
a sample of 1000 simulated absorption spectra.

From the $z=0$ output we extract the pdf of the gas density, which is
shown in Figure \ref{fig1}, this will be implemented in the
semi-analytical (hereafter SA) model. Starting from the linear density
field, generated with the right correlation properties predicted by
linear theory (Subsection
\ref{sam}), we assign each pixel a new non-linear density in such a way
that the resulting non-linear pdf is that extracted from
hydro-dynamical simulations.

The gas density-temperature relation is shown in panel $(a)$ of Figure
\ref{fig2}, where we report the mean value in each density bin and the
scatter.  This result is in reasonable agreement with the recent
results obtained by Yoshida {\it et al.} (2002 - their Figure 2) and
also by Dav\'e {\it et al.} (1999), who studied the gas cooling processes
in the framework of Smoothed Particle Hydro-dynamics (SPH)
simulations. We note that the scatter in temperature can be very high
for values of $\log(1+\delta)>-1$, spanning more than one order of
magnitude.  We report the gas density-metallicity relation in panel
$(b)$. The trend here is as expected: low density regions are the most
metal poor, while at large densities the metallicity can be solar.  In
this plot the scatter is extremely large for the intermediate density
regime, i.e. $-1<\log(1+\delta)<0.5$, while it is significantly
smaller for larger overdensities.  This plot has to be compared with
similar plots shown in Cen \& Ostriker (1999b), but we find that here
the scatter is somewhat larger than in previous simulations.

These curves, with the relative scatter, assumed to be drawn from a Gaussian
distribution, will be implemented into the SA model.

\subsection{Simulation of absorption features in X-ray spectra of AGN pairs}
\label{sam}
We start here from the model introduced by Bi and collaborators (Bi
{\it et al.} 1992, 1995; Bi 1993; Bi \& Davidsen 1997 - hereafter, BD97) 
for generating a Ly$\alpha$ absorption spectrum along a line-of-sight. 
This simple model predicts many properties of the absorption lines,
including the column density distribution and the distribution of line
widths ($b$ parameters), which can be directly compared with
observations. 

The BD97 model is based on the assumption that the low-column density
Ly$\alpha$ forest is produced by smooth fluctuations in the
intergalactic medium which arise as a result of gravitational growth of
perturbations. 
Linear density perturbations in the intergalactic medium $\delta_0^{\rm
IGM}({\bf x}, z)$ are related to the underlying linear dark-matter
(DM) perturbations by a convolution, which models the effects of gas
pressure. In Fourier space one has: $ \delta_0^{\rm IGM} ({\bf k}, z) =
W_{\rm IGM}(k,z) D_+(z) \delta_0^{\rm DM}({\bf k})$, where $D_+(z)$ is
the growing mode of density perturbations (normalised so that
$D_+(0)=1$) and $\delta_0^{\rm DM}({\bf k})$ is the Fourier transformed
DM linear over-density at $z=0$. A commonly adopted low-pass filter 
is $W_{\rm IGM}(k,z) = (1 + k^2/k_J^2)^{-1}$, which  
depends on the comoving Jeans wave-length $\lambda_J(z) \equiv 2\pi H_0^{-1} 
\left[ {2 \gamma k_B T_m(z) \over 3 \mu m_p \Omega_{0m} (1 + z)}\right]^{1/2}$,
where $k_B$ is Boltzmann's constant, $T_m$ the gas temperature,
$\mu$ the molecular weight and $\gamma$ the ratio of specific heats.
$\Omega_{0m}$ is the present-day matter density parameter. 
BD97 adopt a simple lognormal transformation to obtain the IGM density
in the mildly non-linear regime from the linear one.

Several approximations are involved in this modelling.  A first point
which can be important is that the Jeans length depends on density and temperature,
and therefore Jeans smoothing should be an adaptive smoothing of the
density field, while the Fourier implementation cannot take into
account this local aspect (Viel {\it et al.}  2002b). In addition, a
filtering of the dark-matter density field at the Jeans scale may be
not very accurate: different filtering functions result in very
different gas density fields and the real `filtering scale' which
reproduces better the gas distribution, in the linear regime, depends
on the thermal history of the IGM (Hui \& Gnedin 1998; Matarrese \&
Mohayaee 2002; Gnedin {\it et al.} 2002).

The modelling of the gas component using the lognormal transformation
and a smoothing at the Jeans length would determine a poor description
of the gas distribution at $z\sim 0$, where the density fluctuations
are in the non-linear regime.  Even the tight relation between density
and temperature for the gas responsible of Ly$\alpha$ absorption is no
longer valid for the warm-hot intergalactic medium. In this case,
hydro-simulations suggest that the scatter in the $T-\delta$ relation,
for the WHIM, can be very large (Dav\'e {\it et al.} 1999; Yoshida {\it
et al.} 2002).

At least three inputs, taken from hydro-dynamical simulations, can be
used to improve the model: {\it{ i)}} the probability distribution
function (pdf) of the gas density; {\it{ ii)}} the gas
density-temperature relation; {\it{ iii)}} the gas density-metallicity
relation.  Viel {\it et al.} (2002b) showed that by using directly the
pdf of the gas density extracted from the hydro-dynamical simulations,
the pdf of the flux predicted by the semi-analytical model is in
significant better agreement with that in 
the hydro-simulations. The other two inputs will allow to take into
account the physical state (temperature and metallicity) of the
structures responsible for the absorptions.

Here we will follow closely the prescriptions used in Viel {\it et al.}
(2002a, 2002b). The modelling is based on the generation in Fourier
space of 1D density and velocity fields, with the right correlation
properties predicted by the theory.

Starting from a linear power spectrum $P(k)$ we generate 1D LOS linear
random fields for the IGM density and peculiar velocity, with the
low-pass filter window function $W_{IGM}$.  If we draw a
LOS in the $x_{\parallel}$ direction at a fixed coordinate $x_{\perp}$
the Fourier component at redshift $z$ of the linear density contrast
and of the linear peculiar velocity field of the intergalactic medium
(IGM) along the LOS are: \ba \delta^{\rm
IGM}_{0\parallel}(k_\parallel,z |{\bf x}_\perp) &=& D_+(z) \int {d^2
k_\perp \over (2\pi)^2} e^{i {\bf k}_\perp \cdot {\bf x}_\perp} W_{\rm
IGM}(\sqrt{k_\parallel^2 + k_\perp^2},z) \delta_0^{\rm
DM}(k_\parallel,{\bf k}_\perp) \; ,\\ v_\parallel^{\rm
IGM}(k_\parallel,z |{\bf x}_\perp) &=& i k_\parallel E_+(z) \int {d^2
k_\perp \over (2\pi)^2} e^{i {\bf k}_\perp \cdot {\bf x}_\perp} {1
\over k_\parallel^2 + k_\perp^2} W_{\rm
IGM}(\sqrt{k_\parallel^2+k_\perp^2},z) \delta_0^{\rm DM}(k_\parallel,
{\bf k}_\perp) \;, \ea with $E_+(z) =
H(z)f(\Omega_m,\Omega_\Lambda)D_+(z)/(1+z)$. Here
$f(\Omega_m,\Omega_\Lambda) \equiv - d\ln D_+(z)/d \ln (1+z)$), and
$\Omega_m$ and $\Omega_{\Lambda}$ are the matter and vacuum-energy
contribution to the cosmic density (e.g. Lahav {\it et al.} 1991),
respectively; $H(z)$ is the Hubble parameter at redshift $z$:
$H(z)=H_0\sqrt{\Omega_{0m}(1+z)^3+\Omega_{0r}(1+z)^2+\Omega_{0\Lambda}}$,
with $\Omega_{0r} = 1-\Omega_{0m}-\Omega_{0\Lambda}$.
In our case we use the $\Lambda$CDM parameters of the previous subsection.

We then simulate the corresponding density and peculiar velocity at a
distance $r_{\perp}$ from the first LOS. The correlation properties
between these 4 fields are computed using an algebraic implementation,
as described in Appendix A1.

The linear density field is mapped into the non-linear one using
directly the pdf of the gas distribution obtained from
hydro-simulations. This has been done with a rank-ordering technique
between the linear density field and the non-linear one, which
intrinsically consists in a monotonic and deterministic mapping of
each generated field. To account for the scatter, we assign each pixel
a new temperature and metallicity value by adding to the mean value
(continuous lines of Figure \ref{fig2}) a random number taken from a
Gaussian distribution with the same standard deviation as the 
hydro-simulations.

Peculiar velocities are approximated by those predicted by linear theory,
as the peculiar velocity field is known to keep linear even on scales 
where the density contrast gets mildly non-linear. 
This simplification should not
be critical in simulating the spectra, as the effect of peculiar
velocities results in a shift of the line and in an
alteration of the profile (BD97), which will have small impact on the
recovered column density of the absorber. 

After having simulated the gas distribution, the photoionization code
CLOUDY (Ferland {\it et al.} 1998) is used to compute the ionization
states of metals.  In this analysis, we have considered both the
collisional ionization and the photo-ionization. The ionizing
background consists of the sum of a UV background, arising from AGNs
and galaxies, and a X-ray background, probably produced by a population
of moderately X-ray luminous AGN at $z\sim 2$ (Fabian \& Barcons 1992),
while for the soft X-ray part ($< 1$ keV) a significant contribution
comes from the WHIM itself (Cen {\it et al.} 1995; Phillips {\it et
al.} 2001).

Among the heavy elements that can produce absorption lines in the
X-rays Oxygen is the most abundant one and produces the strongest
lines. A more complete treatment should also include other elements
such as C, N, Ne, Fe and Si but, in our simplified scheme, we will
concentrate only on Oxygen lines. For a full list of resonant
absorption lines we refer to Verner {\it et al.}  (1996).  We simulate
absorption spectra for the two strongest transitions of the ions OVII
($E=0.57$ keV, $\lambda=21.6 \AA$, $f=0.7$) and OVIII ($E=0.65$ keV,
$\lambda=18.97\AA$, $f=0.42$), with $f$ the oscillator strength.

\begin{figure*}
\center
\resizebox{.7\textwidth}{!}{\includegraphics{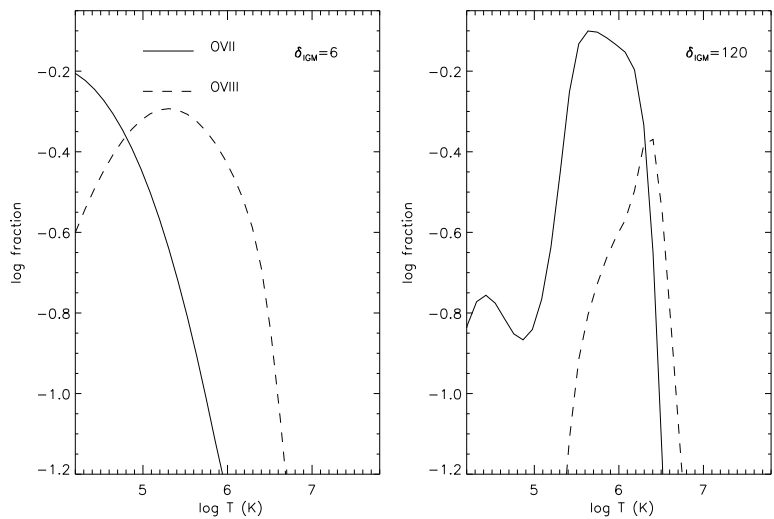}} 
\caption{Logarithm of the ionization fraction as a function of gas
temperature for OVII (continuous line) and OVIII (dashed line). The
left panel shows the results obtained with CLOUDY for an IGM
overdensity $\delta=6$, while in the right panel the results are for an
overdensity $\delta=120$.}
\label{fraction}
\end{figure*}

The density of each ion is obtained through: $n_I(x)= n_H(x) X_I
Y_{Z_{\odot}} Z/Z_{\odot}$, where $X_I$ is the ionization fraction of
the ion as determined by CLOUDY and depends on gas temperature, gas
density and ionizing background, $n_H$ is the density of hydrogen
atoms, Z is the metallicity of the element and $Y_{Z_{\odot}}$ is the
solar abundance of the element.  In Figure \ref{fraction} we plot the
ionization fraction for OVII and OVIII as a function of the temperature
for two different values of the IGM density corresponding to an
overdensity $\delta_{IGM}=6$ (left panel) and $\delta_{IGM}=120$ (right
panel), corresponding to densities of $1.4\times 10^{-6}$ cm$^{-3}$ and
$3\times 10^{-5}$ cm$^{-3}$.  To check if the metals considered here
are in photoionization equilibrium we compute the recombination time
$t_{rec}=(n_{IGM}\, \alpha_{rec})^{-1}$, with $\alpha_{rec}$ the
recombination rate. For OVII and OVIII, in the temperature ranges
considered here, one has that $\alpha_{rec}\sim 2\times 10^{-11}$
cm$^3$ s$^{-1}$ (Mazzotta \et 1998), where both the dielectronic and
radiative recombinations are taken into account. This means that the
Hubble time is larger than the recombination time roughly for densities
$\magcir 10^{-7}$ cm$^{-3}$. Thus, for the WHIM, the approximation of
photoionization equilibrium should be reasonable.

Given the density of a given ion along the LOS, the optical depth 
in redshift-space at
velocity $u$ (in km s$^{-1}$) is 
\be \tau_I(u)= {\sigma_{0,I}c\over H(z)} \int_{-\infty}^{\infty}dy\,
n_{\rm I}(y) {\cal
V}\left[u-y-v_{\parallel}^{\rm I}(y),b(y)\right]
\label{tau} 
\ee 
where $\sigma_{0,I}$ is the cross-section for the resonant absorption
and depends on $\lambda_I$ and $f_I$, $y$ is the real-space coordinate
(in km s$^{-1}$), ${\cal V}$ is the standard Voigt profile normalised
in real-space, $b=(2k_BT/m_I c^2)^{1/2}$ is the thermal width and we
assume that $v_I=v_{IGM}$.  Velocity $v$ and redshift $z$ are related
through $d\lambda/\lambda=dv/c$, where $\lambda=\lambda_I(1+z)$. For
the low column-density systems considered here, the Voigt profile is
well approximated by a Gaussian: ${\cal V}=(\sqrt{\pi}
b)^{-1}\exp[-(u-y-v_{\parallel}^{\rm I}(y))^2/b^2]$.  The X-ray
optical depth $\tau$ will be the sum of the source continuum, which we
assume we can determine, and that of eq. (\ref{tau}). Finally, the
transmitted flux is simply ${\cal F}=\exp(-\tau)$.

\begin{figure*}
\center
\resizebox{.5\textwidth}{!}{\includegraphics{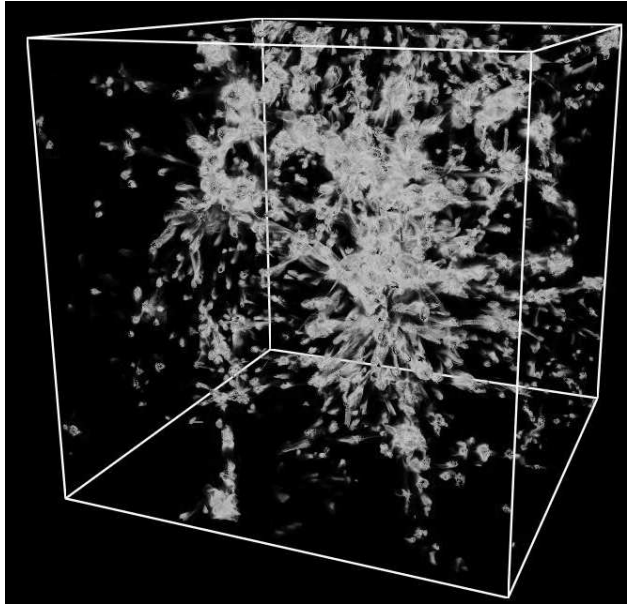}} 
\caption{OVII distribution from the $z=0$ output of the
hydro-simulation ($\Lambda$CDM model, Cen {\it et al.} 2001). OVII
results in a network of diffuse blobs and filaments connecting high
density regions. }
\label{figprova}
\end{figure*}

We generate 400 spectra with our semi-analytical technique at a median
redshift of $z \sim 0.05$ and $\sim 12000$ km/s long.  Both for the
semi-analytical model and for the hydro-simulations spectra have been
produced with a resolution of $\sim 3$ km s$^{-1}$. The goal is to
reproduce number and characteristic size of these absorbers with the
simulation of AGN spectra by a simple semi-analytical recipe, and
address their detectability. Figure
\ref{figprova} shows the OVII distribution at $z=0$ extracted from the 
hydro-simulation. We can see that the OVII connects high density
regions and displays a network of filaments and diffuse blobs. From
this Figure we can argue that strong absorptions arise in virialized
halos, while the weaker ones, possibly detected by future X-ray
missions, sample filaments connecting high density regions.  The OVIII
distribution is quite similar to the OVII one (Chen {\it et al.} 2002).

\section{Testing semi-analytical models with hydro-dynamical simulations}
\label{test}
In this Section we show how well the semi-analytical model works, after
having calibrated it with the inputs obtained from hydro-dynamical
simulations: the pdf of the gas, the gas density-temperature relation
and the gas density-metallicity relation (with the relative scatter)
and the X-ray background. We compare the column density distribution
function for OVII and OVIII predicted by the semi-analytical model with
those extracted directly from the hydro-dynamical simulations.  To this
aim, the SA method has been run cutting off the large scale power for
scales larger than the box-size of the hydro simulation (25$h^{-1}$
Mpc). This allows a more direct comparison between the hydro-simulation
and the semi-analytical model. We checked that the statistics
considered here are only marginally influenced by this assumption, by
running another set of SA spectra without the cut in the
power-spectrum.  We conclude that the box-size of the hydro-simulations
is large enough to study reliably these absorbers.  Column densities
have been computed by integrating directly the simulated optical depth
and not using any fitting routine available, both for the
hydro-simulations and for the SA method. Usually, the number of
absorbers found is small and they can be very easily identified in our
mock-spectra. However, we checked that the estimated column densities
are generally in good agreement with the more accurate results obtained
with fitting routines.

\begin{figure*}
\center
\resizebox{1.\textwidth}{!}{\includegraphics{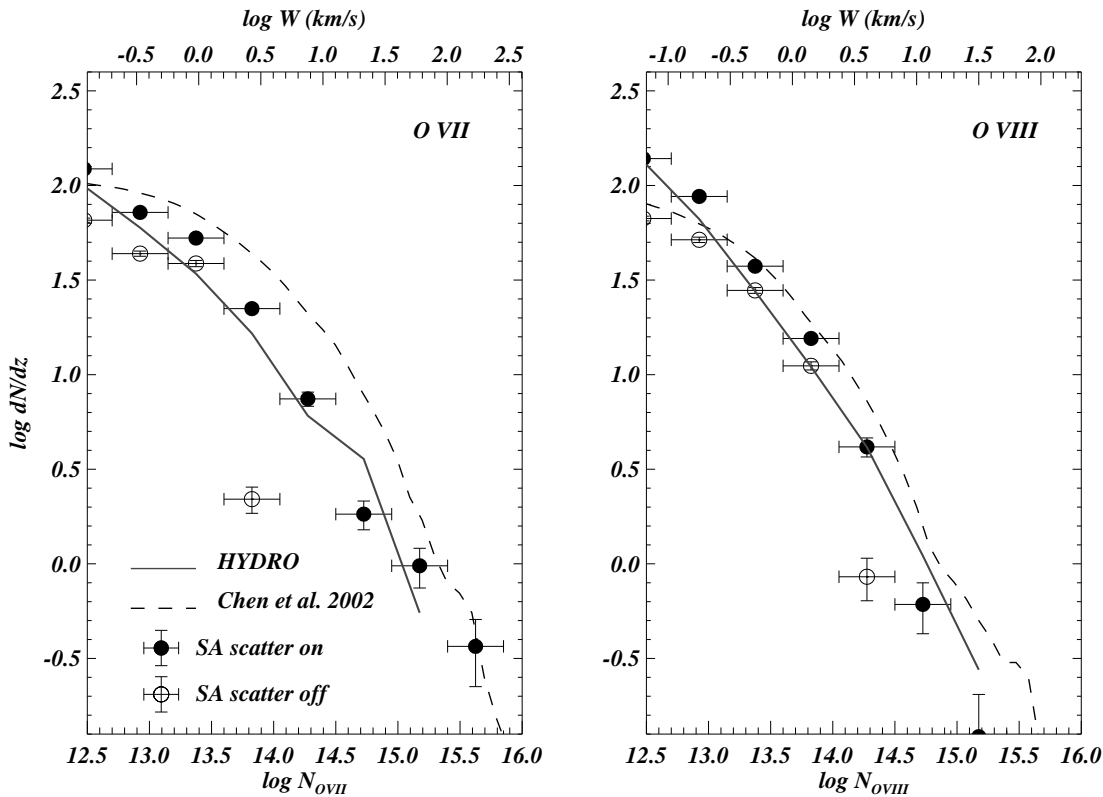}} 
\caption{The left panel shows the cumulative column density
distribution function for OVII (number of absorbers with column density
larger than a given value per unit redshift). Filled circles are from
the semi-analytical model with the scatter (i.e. implementing both the
scatter in the temperature-density relation and in the
density-metallicity relation in the model), empty circles are from the
semi-analytical model without considering the scatter. The continuous
line represents the result obtained from hydro-simulations. The dashed
line show the results of Chen \et (2002), for a model which contains
scatter in the metallicity and a trend of mean metallicity with gas
overdensity: $ \langle\log Z/Z_{\odot} \rangle=-1.66+0.36\log\delta$. 
The right panel shows the same quantities for OVIII. Error bars are 
Poissonian.}
\label{fig3}
\end{figure*}

In Figure \ref{fig3} we show the cumulative number of absorbers per
unit redshift range as a function of their column density (bottom
axis), i.e. the number of systems with column density larger than a
given value, and as a function of the corresponding equivalent width
(top axis), for OVII (left panel) and OVIII (right panel): the
continuous lines represent the results obtained from hydro-dynamical
simulations, while the points are from the SA method. The absorption
equivalent width, in velocity units, is related to the column density
by the simple relation: \be W=\frac{\pi e^2}{(m_e c)} \, \lambda f\, N \ee
where $N$ is the column density, $f$ the oscillator strength and
$\lambda$ the wavelength of the transition (Sarazin 1989, Chen {\it et
al.}  2002). This relation formally holds in the limit of optically
thin absorber but is also a good approximation for column densities
smaller than $10^{15} \rm{cm}^{-2}$.

To quantify the role of the scatter in our simulations, we
decided to run another set of SA spectra and switch-off any form of
scatter, both in the temperature and metallicity. Results without
scatter are shown as empty circles, while those that include the scatter
are represented by filled circles.  From Figure \ref{fig3} we see that,
when the scatter is considered, the agreement between
hydro simulations and SA is good. A first conclusion is that the
inputs we used for the SA method, taken from hydro-simulations, are
enough to have a good description of the number of absorbers per unit
redshift range for the column density range $10^{12.5}$cm$^{-2} <
N_{OVII,OVIII} < 10^{15.5}$cm$^{-2}$, corresponding to $0.1 {\rm km/s} \mincir
\rm{W} \mincir 35$ km/s for $O_{VII}$.

The values found are in agreement with the column densities
distributions of Chen {\it et al.}  (2002), which are reported in
Figure \ref{fig3} and are represented as a dashed line. In the case in
which they take into account the metallicity-density relation,
parameterized by $\langle\log Z/Z_{\odot} \rangle=-1.66+0.36\log\delta$ 
and consider
a scatter drawn randomly from a lognormal distribution with $\langle\log
Z/Z_{\odot} \rangle=-1$ and $\sigma_{\log Z}=0.4$, the results are quite
similar.
We note
a slightly steeper column density distribution function in our analysis
compared to the one in Chen {\it et al.} (2002), which results in a
smaller number of absorbers at intermediate column densities in the
range $10^{13}-10^{15}$
cm$^{-2}$. This is probably determined by the fact that our gas-density
metallicity relation predicts lower metallicities than the one they
used.

Before making any consideration about the detectability of these
systems, we make some comments about the effect of the scatter. If we
do not take into account the scatter, we cannot reproduce all the
systems at column densities larger than $10^{14}$ cm$^{-2}$ for both
the ions.  For smaller column densities the effect of the scatter is
smaller, but it seems that also in this range the scatter produces an
increase in the number of systems.

We checked the different role played by the metallicity and temperature
scatter by running another set of SA spectra switching off only one of
them.  It turns out that the scatter that influences most the column
density of the absorbers is the one in metallicity.
This is probably due to the fact that the column
density of the absorbers is directly proportional to the gas
metallicity (Section \ref{sam}). The dependence on the temperature is
in a sense less direct, as the latter determines the ionization fraction,
computed with CLOUDY including both the collisional ionization and the
photo-ionization contributions (see Section \ref{sam}), which depends
on the particular X-ray background chosen in a non-trivial way (see for
example Figure \ref{fraction}).  Another reason could be
that the scatter in metallicity is slightly larger than the temperature
one for the regions with $5<\delta<10$, a low density WHIM, which is responsible for a significant fraction of the absorptions.

We expect $\magcir 30$ OVII systems with column densities larger than
$10^{13} \rm{cm}^{-2}$ corresponding to equivalent widths larger than
2 km/s, per unit redshift range at $z\sim 0$. While the number of
systems with column densities larger than $10^{14.5} \rm{cm}^{-2}$,
corresponding to equivalent widths larger than $\sim 10$ km/s, drops
to 5.  The number of OVIII systems is similar to that of OVII ones
showing that both ions trace structures at the same density, but the
corresponding OVIII equivalent width is almost a factor 2 lower.

\section{Results}
\label{resu}
\begin{figure*}
\center
\resizebox{1.\textwidth}{!}{\includegraphics{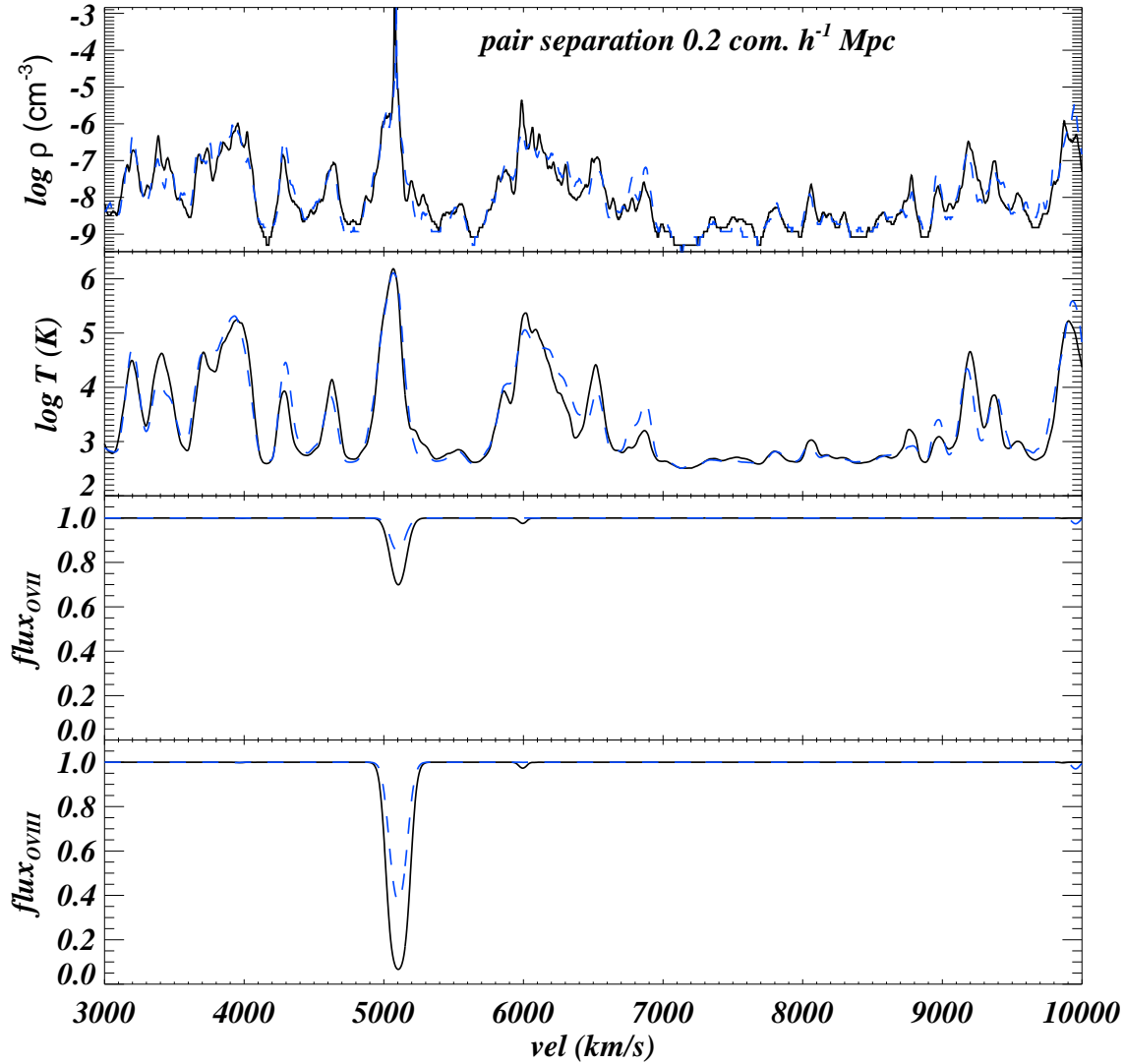}} 
\caption{From top to bottom: gas density along the line of sight,
temperature, spectrum for OVII and spectrum for OVIII. Continuous line
and dashed line represent the two AGN spectra of a pair with 
separation  r$_{\perp}=0.2$ comoving $h^{-1}$ Mpc at $z \sim 0.05$, 
simulated with the
semi-analytical technique described in the text. This portion of the 
spectrum has been selected around one of the strongest absorptions. }
\label{fig4}
\end{figure*}

\begin{figure*}
\center
\resizebox{1.\textwidth}{!}{\includegraphics{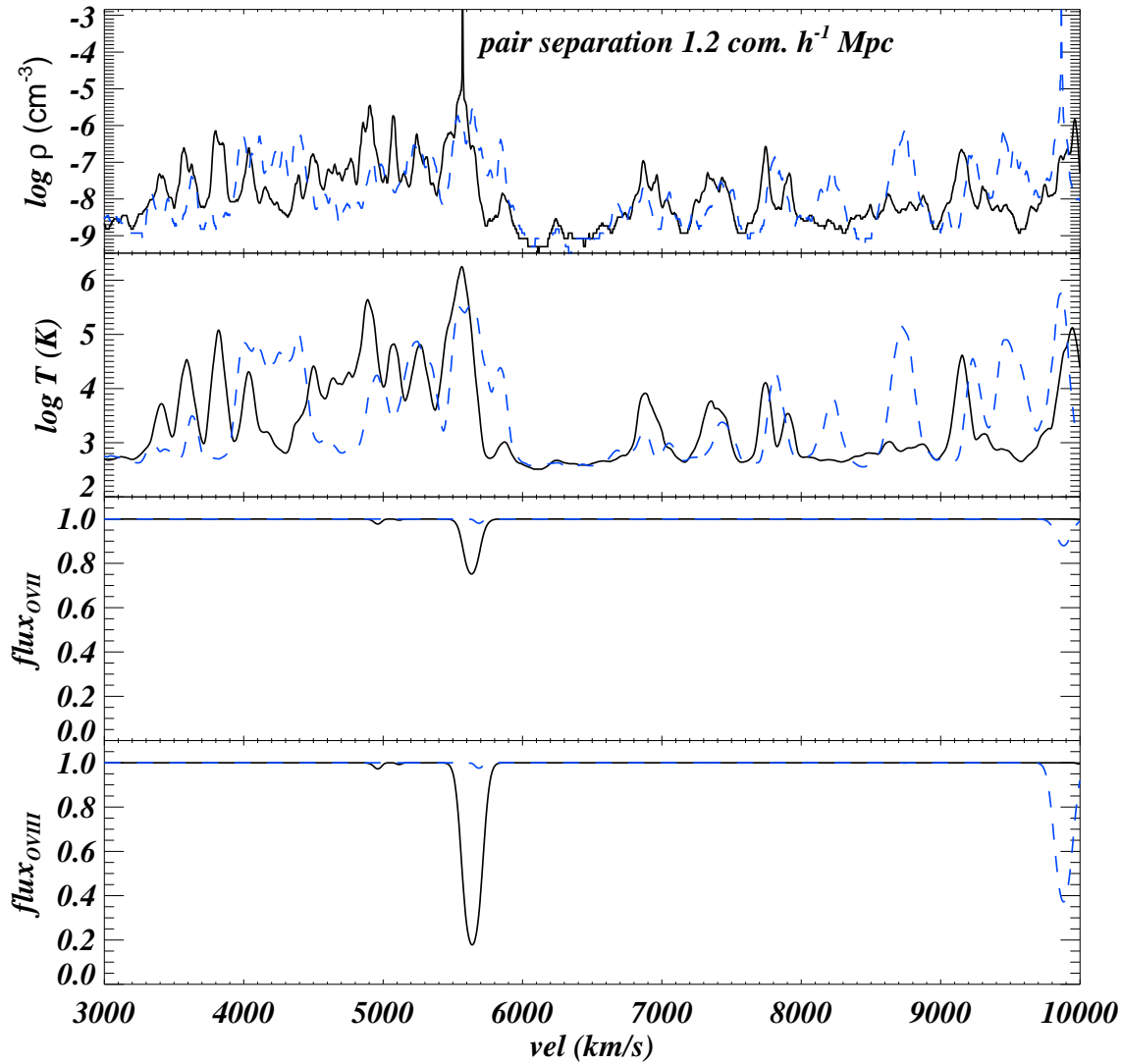}} 
\caption{From top to bottom: gas density along the line of sight,
temperature, spectrum for OVII and spectrum for OVIII. Continuous line
and dashed line represent the two spectra of the AGN pair with a
separation of r$_{\perp}=1.2$ comoving $h^{-1}$ Mpc at $z \sim 0.05$,
simulated with the semi-analytical technique described in the
text. This portion of the spectrum has been selected around one of the
strongest absorptions.}
\label{fig5}
\end{figure*}

To better understand the physical state of the gas which determines the
absorptions we plot in Figure \ref{fig4} and Figure
\ref{fig5} two AGN pairs, for two different separations of 0.2 $h^{-1}$
Mpc and 1.2 comoving $h^{-1}$ Mpc (continuous line and dashed line
represent the two members of the pair) simulated with the
semi-analytical technique. Assuming that the absorbers are placed at a
median redshift of $z \sim 0.05$, these distances correspond to an
angular separation of $\sim 3.5$ arcmin and $\sim 42$ arcmin, respectively,
for a $\Lambda$CDM Universe. If the absorbers are placed at a median
redshift of $z \sim 0.1$ the angular separation is $\sim 2$ arcmin and $\sim
30$ arcmin, respectively.

In Figure \ref{fig4} and \ref{fig5} it is possible to see that the
absorptions arise from gas at a temperature larger than $10^5$ K which
has a density larger than $10^{-6}$ cm$^{-3}$,
i.e. overdensity $\delta  > 6$.
One can see that, while for the smaller separation (Figure \ref{fig4}) the
two AGNs show almost identical features both in density and  
temperature, which determine very similar spectra for OVII and OVIII,
for the largest separation of Figure \ref{fig5} the differences are
larger. This is due to the fact that now the correlations are
weaker, resulting in more different density and temperature profiles
for the two AGN spectra. Also the spectra now show some coincident
absorptions, for example at 5600 km s$^{-1}$, and some anticoincidences,
such as the strong one at $\sim 10000$ km s$^{-1}$, while at smaller
separations there are only coincident features (Figure \ref{fig4}).

To quantify the amount of correlation between the spectra of the AGNs
of the simulated pairs we use the 'hits-and-misses' statistics
described by McGill (1990), Crotts \et (1994), Bechtold \et (1994),
Charlton {\it et al.}  (1997).  A coincidence is defined as the case in
which an absorption line is present in both the spectra within a given
velocity difference $\Delta v$ and above some signal-to-noise ratio. An
anticoincidence is defined when a line is present in one spectrum but
not in the other one.  If there are two lines within $\Delta v$, a
pretty rare event given the incidence of OVII and OVIII absorbers, we
count only one coincidence and no anticoincidence, as in Fang {\it et
al.} (1996).  We also generate a set of spectra uncorrelated in the
transverse direction and compute the
same statistics to take into account the level of random coincidences.
Despite the arbitrariness of the definition, which involves the choice
of a velocity binning for the spectra, it has been shown that it is
still useful to define a characteristic size for the observed
Ly$\alpha$ absorbers (Charlton {\it et al.}  1997; Fang {\it et al.}
1996). The inferred 'characteristic size' of the Ly$\alpha$ clouds is
in agreement with the results of other hydro-dynamical simulations
(Theuns {\it et al.} 1998) and with semi-analytical models of the
Ly$\alpha$ forest (Viel {\it et al.}  2002a, Viel {\it et al.} 2002b).

\begin{figure*}
\center
\resizebox{1.\textwidth}{!}{\includegraphics{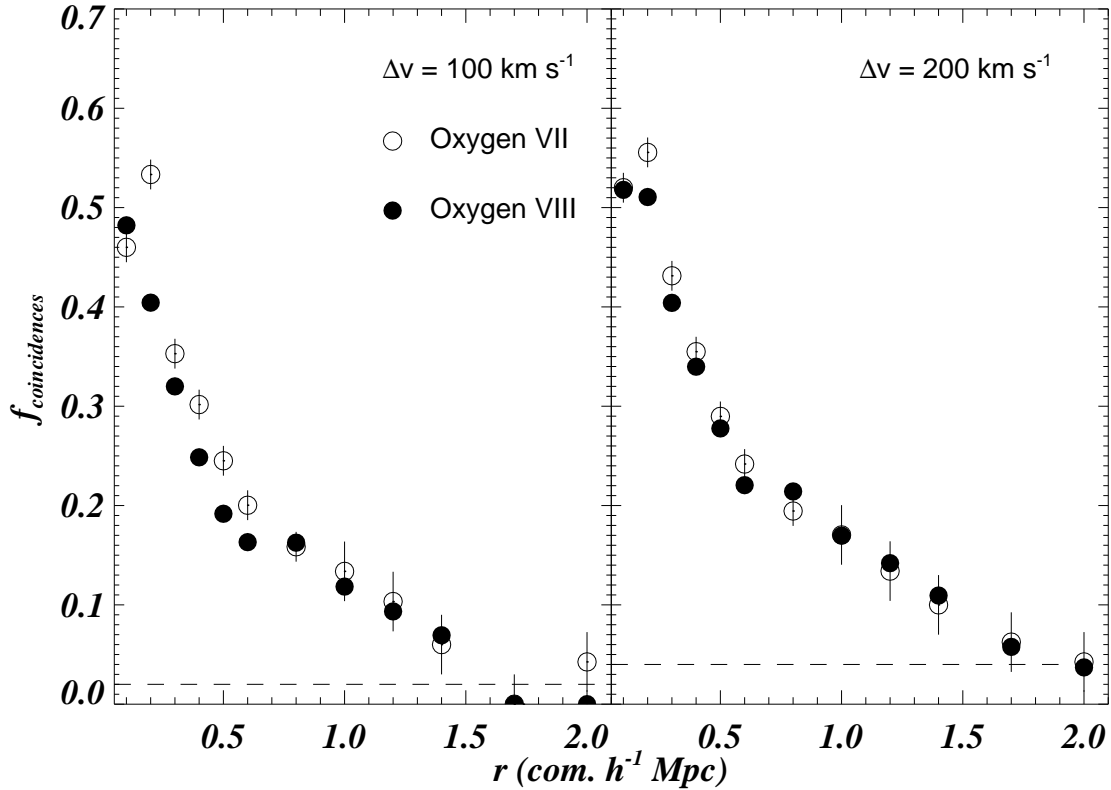}} 
\caption{The coefficient representing the fraction of coincidences, as
defined in the text, i.e. the number of 'hits' divided by the total
number of 'hits' and 'misses', plotted as a function of the separation
between the two AGNs. Only absorbers with column densities larger than
$10^{13}$ cm$^{-2}$, which can be detected by {\it XEUS} and {\it
Constellation-X}, have been selected to compute this coefficient. The
chosen bin to compute the hits and misses statistics is $\Delta v=100$
km s$^{-1}$ in the left panel and $\Delta v=200$ km s$^{-1}$ in the
right one. Empty circles represent results for OVII and filled ones for
OVIII. The dashed line is the level of random coincidences, which depends
on the velocity bin chosen. The points have been obtained by simulating 20
pairs for each separation. Error bars (plotted only for OVII)
represent the errors of the mean.}
\label{fig6}
\end{figure*}

Results are shown in Figure \ref{fig6} where we plot the coefficient
$f_{co}$, the fraction of coincidences, which is defined as the ratio
between the number of 'hits' and the total number of 'hits' and
'misses': the larger the value of this coefficient the larger the
correlation between the two spectra. These plots have been
obtained by simulating absorption spectra of 20 AGN pairs at a median
redshift of $z\sim 0.05$ for 12 different separations, spanning the
range $0.1 - 2$ comoving $h^{-1}$ Mpc, that is angular separations from 3
to 70 arcmin. OVII and OVIII absorbers are represented by empty and
filled cirlces, respectively. This coefficient has been computed {\it
only for absorbers with column densities $> 10^{13} \rm{cm}^{-2}$},
which will be detectable by {\it XEUS} and {\it
Constellation-X}.  We have decided to show this result selecting all
the absorbers detectable by future X-ray missions, i.e. all
the absorbers with column densities larger than $10^{13}
\rm{cm}^{-2}$. This assumption is not crucial: the values assumed by
the coefficient $f_{co}$ are pretty robust and change significantly
only if we include the absorbers with lower column densities. In fact,
for these absorbers the level of random coincidences is higher and
determines higher values of $f_{co}$. The values assumed by $f_{co}$
are mainly determined by the absorbers with column densities in the
range $10^{13.5}-10^{15}$ cm$^{-2}$. For larger column densities
$f_{co}$ becomes smaller because the number of random coincidences
found is smaller.

In the left panel a bin in velocity of 100 km s$^{-1}$ has been chosen
for the hits-and-misses statistics, while in the right one the bin is
of 200 km s$^{-1}$. The dashed line is the level of random coincidences
calculated from the 'hits-and-misses' analysis of fully uncorrelated
AGN spectra. For both absorbers the coefficient peaks within $\sim
0.2 $ comoving $h^{-1}$ Mpc and then drops. In both panels the value at
which the coefficient becomes comparable to the random coincidence
level is at $\sim 1.2$ comoving $h^{-1}$ Mpc. Of course the number of
random coincidences depends on the binning chosen, being smaller for
smaller $\Delta v$. However, we stress that the number of random
coincidences is very small, given the small number of systems expected
per unit redshift path.
 
Assuming that the noise-level for $f_{co}$ is of the order of 0.05, as
we can infer from the right panel of Figure \ref{fig6}, a 3$\sigma$
detection of this filamentary structure will require at least 20 pairs
with separations $\mincir 1$ comoving $h^{-1}$ Mpc corresponding to the
range $\mincir 20$ arcmin if the absorbers are placed at $z \mincir
0.1$. Another conclusion is that a sample of at least 20 AGN pairs
with this separation is necessary to discriminate between random
coincidences and 'real' signal from this filamentary structure.

\subsection{{\it XEUS} and {\it Constellation-X} detectability of OVII 
absorbers}

In this Section we quantify the detectability of OVII absorbers,
following the treatment and formalism of Sarazin (1989).  In
particular, we compute detection thresholds of the filamentary network
for single AGN spectra and for AGN pairs. This means that in the
latter case we will need 20 pairs of sources with separation $\mincir
20$ arcmin, in a given portion of the sky, to probe the characteristic
size of the absorbers by analysing coincident absorption features (see
Section \ref{resu}).  We decided to focus only on OVII absorbers. As
stressed above OVII and OVIII absorbers probe the same structures,
but, for a given column density the corresponding OVII absorber
equivalent width is a factor 2 higher than the OVIII one. This means
that OVII absorbers can probably be detected in slightly lower density
regions than OVIII ones, so they can better trace the bulk of the
filamentary structure.

In the following discussion  all the fluxes are expressed in the 
ROSAT All-Sky Survey (RASS) 0.1-2.4 keV energy band. 

In order to detect absorption lines, with a specified signal-to-noise
level $S/N$, the detector must collect, within the instrument energy
resolution $\Delta E$, a number of continuum photons given by
$N_{photons}\magcir \left(\frac{S}{N}\right)^2\left(\frac{W_{abs}}
{\Delta E}\right)^{-2}$, where $W_{abs}$ is the absorber equivalent
width.
The corresponding flux at the line energy E is: 
\be 
F\,\magcir 1.3 \times
10^{-12}\left(\frac{E}{1\,\rm{keV}}\right)^{-1}\left(\frac{S/N}{3}
\right)^2\left(\frac{\Delta
E}{10\, \rm{eV}}\right) \left(\frac{W_{abs}}{100\, \rm{km/s}}\right)^{-
2}
\left(\frac{A\,Q}{10000\,\rm{cm}^2}\right)^{-1}\left(\frac{t}{100\,
\rm{ksec}}\right)^{-1} \rm{erg\, cm}^{-2}\, \rm{s}^{-1}\,
\rm{keV}^{-1} \, ,
\label{flux}
\ee 
where $AQ$ is the effective collecting area ($Q$ is the quantum 
efficiency of the detector) and $t$ is the integration time.

We estimate the source flux needed to detect OVII absorption lines
($E=0.57$ keV, in the observer rest frame) using different X-ray
telescopes and exposure times of 500~ksec.  Although a conservative
line detection requires a $S/N=5$, in all our calculations below we
choose $S/N=3$.  

%This is justified by the fact that successive
%follow-up at different wavelengths (for example searching the
%corresponding OVI absorption lines in the UV band) may later be used to
%confirm or not the presence of the absorbers (see e.g. Mathur {\it et
%al.}  2002).

We find that the detection of these absorbers is quite
difficult.  The resolution and collecting areas at $\sim 0.5$ keV of
the present generation X-ray satellites are: $R= E/\Delta E= 500$,
$AQ=25$~cm$^2$ and $R=500$ and $AQ=70$ ~cm$^2$ for {\it Chandra} and {\it
XMM-Newton}, respectively.  If we substitute these values in
eq. (\ref{flux}) we find that column densities of $10^{14}$
cm$^{-2}$, corresponding to equivalent widths $W\mincir 10$ km s$^{-1}$,
cannot in practice be detected by these satellites: they either
require exceptionally bright background sources with fluxes $\sim
10^{-9} \rm{erg\, cm}^{-2}\, \rm{s}^{-1}$ (calculated assuming a
source spectrum F $\propto E^{-0.7}$) or
unrealistic exposure time of the order of the {\it Chandra} life-time.  On the other hand,
if we consider more realistic, although still quite bright and rare,
X-ray sources with $F \sim 10^{-11} \rm{erg\, cm}^{-2}\, \rm{s}^{-1}$ we
find that the only observable absorption features detectable by {\it
Chandra} and {\it XMM-Newton} are the ones with $W > 200$~km/s (see
also Mathur {\it et al.} 2002; Fang {\it et al.} 2002b).

Conversely, the low density network of filaments can be detectable by
next generation satellites, with much higher collecting area, such as
{\it Constellation-X} {\footnote{http://constellation.gsfc.nasa.gov}}
and especially {\it XEUS}
{\footnote{http://astro.estec.esa.nl/SA-general/Projects/XEUS}. For
these satellites the effective collecting area and resolution are:
$R=1500$, $AQ=3000$ cm$^2$ for {\it Constellation-X} and $R=800$ and
$AQ=40000$ cm$^2$ for {\it XEUS}. We mention here that the value
$R=1500$ for constellation-X is that expected from the off-plane option
for the Reflection Grating Spectrometer (W. Cash 2002,
http://constellation.gsfc.nasa.gov) and that for {\it XEUS} the
ultimate satellite configuration {\it XEUS-2} will allow to reach an
effective area of $AQ=2\times 10^5$ cm$^{2}$.

Assuming that sources with a flux $\sim 10^{-11} \rm{erg\,
cm}^{-2}\, \rm{s}^{-1}$ are observed, we find that the minimum detectable 
equivalent width is $\sim 6$ km s$^{-1}$ and $\sim 3$ km s$^{-1}$, for {\it
Constellation-X} and {\it XEUS}, respectively. With {\it XEUS-2} the
minimum detectable equivalent width will be $\sim 1.5$
km~s$^{-1}$. These equivalent widths will probe the range of absorbers
of column densities of $\magcir 10^{14.6}$ cm$^{-2}$ for {\it
Constellation-X}, $\magcir 10^{13.8}$ cm$^{-2}$ for {\it XEUS} and
$\magcir 10^{13.5}$ cm$^{-2}$ for {\it XEUS-2}.

We now estimate the number of AGN with fluxes sufficient to allow
detection of absorbers of given $W_{abs}$. We use the fit to the
log N - log S relation ${\cal N}(>S)\sim 5.1\times 10^{-16} (S/\rm{erg}\ {\rm
cm}^{-2}$ {\rm s}$^{-1})^{-1.5} {\rm{sr}}^{-1}$, originally calculated in the
{\it Einstein} band and here renormalized to the {\it ROSAT} band (see
Maccacaro {\it et al.} 1982; Hasinger {\it et al.} 1993).  Formally
this fit is valid for fluxes in the range $10^{-13} - 10^{-11}
\rm{erg\, cm}^{-2}\, \rm{s}^{-1}$ and extrapolation to lower fluxes
can be dangerous as we know, for example, that the ROSAT Log N - Log S
function in the Lockman Hole has a shallower slope than the one used 
here at lower fluxes (Hasinger {\it et al.} 1998).  Nevertheless, as
we are interested in very bright sources, the Maccacaro {\it et al.} fit
should be a good approximation.  Assuming that the typical quasar
spectrum is given by $F \propto E^{-0.7} $ from eq. (\ref{flux}) and the
adopted Log N - Log S relation we find: \be {\cal N}(>F) \approx 100
\left(\frac{E}{1\,\rm{keV}}\right)^{0.45}\left(\frac
{S/N}{3}\right)^{-3}\left(\frac{\Delta E}{10\, \rm{eV}}\right)^{-1.5}
\left(\frac{W_{abs}}{100\, \rm{km/s}} \right)^{3}
\left(\frac{A\,Q}{10000\,\rm{cm}^2}\right)^{1.5}\left(\frac{t}{100\,
\rm{ksec}}\right)^{1.5} \rm{sr}^{-1} \, .  \ee

\begin{figure*}
\center
\resizebox{.5\textwidth}{!}{\includegraphics{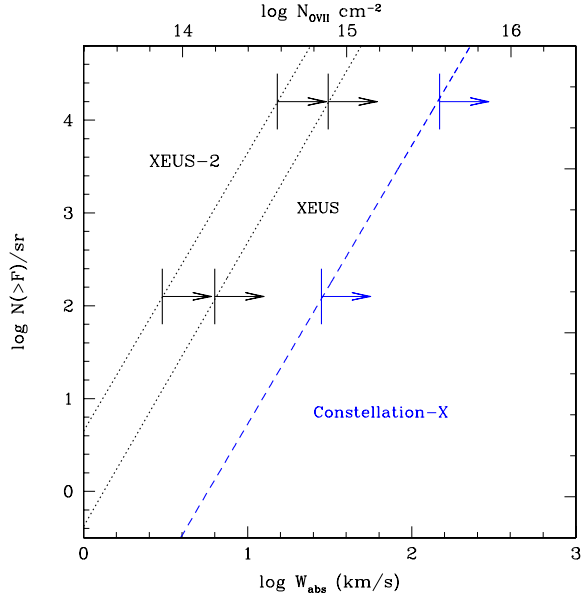}} 
\caption{Cumulative source number per steradian with enough
flux to allow detection of OVII absorption lines with a given
equivalent width. The continuous line is for {\it XEUS}, the dotted one for
{\it XEUS-2} and the dashed one for {\it Constellation-X}, respectively.
We assumed S/N=3 and exposure times of 500~ksec.  Vertical lines mark
thresholds for which the structures can be probed by AGN pairs. The
first threshold on the left represents the estimated minimum W
detectable for OVII from a sample of 20 pairs, with separation
$\mincir 20$ arcmin, from an observation of the whole sky. The second
threshold represents the same value but obtained from a sample of 20
pairs found in a smaller random field of 4 deg$^2$. In the upper axis
the corresponding OVII column densities are reported. }
\label{fig7}
\end{figure*}

In Figure \ref{fig7} we plot the number of sources per steradian
with enough flux to allow the detection of absorbers of a given
equivalent width. The solid line refers to {\it XEUS}, the dotted one
to {\it XEUS-2} and the dashed one on the right refers to {\it
Constellation-X}.  In the same Figure we also report the corresponding
OVII column densities.

From Figure \ref{fig7} one can see that {\it XEUS} can detect roughly
10 sources per steradian with equivalent width $\sim 2.5$ km/s,
corresponding to column densities of $10^{13.8}$ cm$^{-2}$ for
OVII. For these sources, at least a number of 30 features can be seen
in absorption per unit redshift range if the current models for the
WHIM are correct (Figure \ref{fig5}). {\it XEUS-2} will find a factor
of 10 more sources per steradian than {\it XEUS}, which will allow to
probe the same WHIM structures. {\it
Constellation-X} will not be able to probe the same very low density
range. For example, in this case 10 single sources per unit solid
angle, bright enough to probe absorbers with an equivalent width of
$\sim 10$ km s$^{-1}$, are predicted to be in the sky.  The number of
absorption features with this equivalent width is approximately 1 per
unit redshift range.

We estimate now the  number of AGN pairs with flux above a given value
and whose relative separation is between $\theta_{min}$ and $\theta_
{max}$.
To this aim  we consider a survey area  $\Sigma$ and we use the  
 X-ray sources angular correlation function 
$w(\theta)=(\theta/\theta_0)^{-0.8}$, with
$\theta_0 \sim 3$ arcmin  (Akylas {\it et al.}  2000).

If we indicate with ${\cal N}_{\Sigma}=\int_{\Sigma} {\cal N}(>F) \, 
d\Omega$ the total
number of sources in the survey with flux above a given value,
then the number of distinct pairs whose members
have a relative separation between $\theta_{min}$ and $\theta_{max}$ 
is: 
\be {\cal N}_{pairs}(\theta_{min},\theta_{max})= \pi \, {\cal N} \, {\cal N}_{\Sigma}
\int_{\theta_{min}}^{\theta_{max}}(1+w(\theta))\sin\theta d\theta \approx \frac{\pi\,{\cal N} \, {\cal N}_{\Sigma}}{2}\,[{\theta_{max}^2}+1.7\,\theta_0^{0.8}\,\theta_{max}^{1.2}]\; ,
\label{npairs}
\ee 
where the last expression is valid for small separations and for
$\theta_{min}=0$.

We choose the limits $\theta_{min} = 0$ and $\theta_{max} =
20$ arcmin, in computing the number of pairs. 
For these angular separations, as we have found in Section \ref{resu},
we expect to find significant coincidence features in the spectra of AGN pairs.

The vertical bars in Figure \ref{fig7} represent the minimum absorber
equivalent width which can be probed by using AGN pairs, instead of
single AGN spectra.  Detection of the filamentary structure with future
missions can be achieved by finding coincident absorptions in AGN
pairs, as we showed in Section \ref{resu}.

The first threshold on the left in
Figure \ref{fig7} corresponds to the equivalent width which can be
probed from a sample of 20 pairs, with the right separations, found in
the whole sky, while the threshold to the right corresponds to a field
of 4 deg$^2$.  This means that, if {\it XEUS} or {\it Constellation-X}
will observe a random field of 4 deg$^2$ in the sky, they will find 20
AGN pairs with the wanted separation with fluxes high enough to probe
the WHIM structures at the corresponding equivalent width.  
If this is the case, the expected number of OVII absorption features
seen by XEUS per unit redshift range is $\sim 2.5$. Thus, if these
objects are placed at $z \sim 0.5$ we predict that a sample of 20
spectra of AGN pairs will possibly show a total of $\sim 50$
absorption OVII features, while {\it Constellation-X} will detect a
factor of 3 less absorption features. The more powerful {\it XEUS-2}
will detect $\sim 100$ absorptions in the same
conditions. An observation of the whole sky will possibly allow these
satellites to detect WHIM structures at smaller column densities.

To conclude we verify if the {\it ROSAT} all-sky survey bright source
catalogue (Voges {\it {et al.}} 1999) may contain the positions of a
sufficient number of pairs to probe the WHIM structure. This catalogue
contains a sample of 18,811 sources down to a limiting {\it ROSAT} PSPC
count- rate of 0.05 cts s$^{-1}$ in the 0.1 - 2.4 keV band, which
corresponds to a flux limit $5.2 \times 10^{-13}$ erg cm$^{-2}$
s$^{-1}$ (assuming a photon spectrum which can be modeled by a power
law with slope -1 and no absorption, 1 cts/s $\sim 1\times
10^{-11}$ erg cm$^{-2}$ s$^{-1}$). We find that the number of pairs of
sources whose separation is smaller than 20 arcmin is $\sim 100$ for
fluxes larger than $3.3 \times 10^{-12}$ erg cm$^{-2}$ s$^{-1}$, and
is $\sim 300$ for fluxes larger than $2.2 \times 10^{-12}$ erg cm$^2$
s$^{-1}$.  If we consider that approximately 16 \% of the X-ray
sources have unique counterpart of extragalactic origin (Voges {\it
{et al.}} 1999), we deduce that the RASS catalogue should contain the
positions of $\sim 48$ pairs of extra-galactic sources with fluxes
$\magcir 2 \times 10^{-12}$ erg cm $^{-2}$ s$^{-1}$. These bright
sources will allow {\it XEUS} to probe absorbers of $W \sim 7$ km
s$^{-1}$ (see eq. (\ref{flux})).

\section{Discussion and conclusions}

In this paper we have presented a semi-analytical technique to simulate
X-ray absorption spectra of distant AGNs.  The Warm Hot Intergalactic
Medium (WHIM) can be detected both in absorption and in
emission. Recently, a number of  
missions devoted to the study of the WHIM through the observation
of emission lines (see e.g. the missions {\it SPIDR}
{\footnote{http://www.bu.edu/spidr/} or {\it IMBOSS}
{\footnote{http://www.ias.rm.cnr.it/IMXS/iasimxs.html}) 
have been proposed.  Unlike the
line emission process, which is proportional to the density square, the
line absorption has the advantage of being proportional to the density of
the intervening absorber.  Furthermore, absorption measurements are not
subjected to background contamination.  Absorption measurements, then,
happen to be more sensitive than emission ones to the lower density
regions of the WHIM. Having a larger volume filling factor, these low
regions constitute the bulk of the filamentary structure.  Thus, 
absorption measurements, besides representing an important tool 
complementary to emission ones, are probably more promising to study 
the large scale structures traced by the WHIM.

The method proposed here uses as input four results obtained from the
$z=0$ output of a large box-size $\Lambda$CDM hydro-dynamical
simulations: the probability distribution function of the gas, the gas
temperature-density relation, the gas metallicity-density
relation. These quantities are sufficient to give a good description of
the number of OVII and OVIII absorbers expected in the low redshift
Universe. Simulations of these two ions are very important: they are
the strongest trace of the WHIM.  If we sum up all the baryons at
$z\sim 2$ and compare this value with the baryons seen at $z\sim 0$,
we discover that a significant fraction of the baryons is missing (Cen
\& Ostriker 1999). This fraction is thought to be in the WHIM. Thus,
the detectability of this network of OVII and OVIII filaments is
of fundamental importance. 

We have extended the simulation of a line-of-sight (LOS) to the
simulation of pairs of LOS with the mathematical implementation reported
in Appendix A1 (see also Viel {\it et al.} 2002a). This has been done 
in order
to give an estimate of the characteristic size of the absorbers by
using the information contained in the transverse direction. While
single LOS are sensitive to the size of single absorbers, pairs of LOS
should be more reliable a probe of the network of filaments traced by
the metals as they are more sensitive to the random orientation of
each filament. We have performed the so called `hits-and-misses'
statistics (McGill 1990; Charlton {\it et al.} 1997) on a sample of
simulated pairs to obtain an estimate of this size.

The main result of the first part of the paper can be summarized as
follows: {\it i)} semi-analytical models successfully reproduce the
number of absorbers per unit redshift path of hydro-dynamical
simulations and we expect $\magcir 30$ absorbers per unit redshift
with an OVII column density $>10^{13.5}$ cm$^{-2}$; {\it ii)} scatter
in the metallicity-gas density relation plays a significant role in
increasing the number of absorbers with column densities $\magcir
10^{14}$ cm$^{-2}$; {\it iii)} the characteristic size of these
absorbers at $z\sim 0$ obtained by analysing coincident absorptions in
simulated spectra of AGN pairs, is $\sim 1$. 

In the second part we have discussed the detectability of the absorbers
by present X-ray missions, such as {\it Chandra} and {\it XMM-Newton}
and future ones, such as {\it XEUS} and {\it Constellation-X}.  Up to
now only few `trees' of the `X-ray forest' have been detected by {\it
Chandra} (Fang {\it et al.} 2002b, Nicastro {\it et al.} 2002a). Most
likely, these absorbers sample the high column density tail of the
distribution shown in Figure \ref{fig3}. A detection of the lower
column density absorbers, whose volume filling factor is larger, can be
achieved only by future X-ray missions whose collecting area and
sensitivity will be higher.

We have found that at least a background of 20 pairs of bright
extragalactic AGN-like sources at a median redshift of $z\sim 0.5$ and
with relative separations lower than 20 arcmin is needed to probe this
filamentary structures with `hits-and-misses' statistics. We have made
a theoretical estimate to address the detectability of these sources by
{\it XEUS} and {\it Constellation-X} taking into account the bright
X-ray sources angular correlation function (Akylas {\it et al.} 2000).
The main conclusions are: {\it i)} by observing a random field of 4
deg$^2$ {\it XEUS} and {\it Constellation-X} will be able to probe WHIM
absorbers, with `hits-and-misses' statistics, which will correspond to
OVII column densities $10^{14.8}$ cm$^{-2}$ and $10^{15.5}$ cm$^{-2}$,
respectively; {\it ii)} in this field {\it XEUS} and {\it XEUS-2} will
possibly find $\sim 50$ and $\sim 80$ absorption features,
respectively, if we assume that the sources are placed at a median
redshift of $z\sim 0.5$; {\it iii)} {\it XEUS} will detect at least 10
sources per steradian bright enough to probe structures with an
absorber equivalent width of 2.5 km/s, corresponding to an OVII column
density of $10^{13.8}$ cm$^{-2}$; {\it iv)} {\it Constellation-X} will
detect few thousand sources per steradian bright enough to probe
structure with an equivalent width of $\sim 100$ km/s.  We have checked in
the {\it ROSAT} all-sky survey bright source catalogue (Voges {\it et
al.}  1999) to determine if these sources had already been
detected. We have found that the required number of extragalactic sources
with the necessary angular separation is present in the
catalogue. Re-observation of these sources will allow future X-ray
missions to detect baryons in the low redshift Universe and to unveil
the filamentary structure where they preferentially reside.

\section*{Acknowledgements.}
We thank M. Haehnelt, V. Mainieri, G. Matt, M. Mendez, G. Perola,
C. Porciani for useful discussions and technical help. We thank the
anonymous referee for helpful comments and X. Chen for kindly providing
his data. MV acknowledges partial financial support from an EARA Marie
Curie Fellowship under contract HPMT-CT-2000-00132. This work has been
partially supported by the European Community Research and Training
Network `The Physics of the Intergalactic Medium' and by grant
ASC97-40300.

\appendix
\label{algebra}
\section{Correlation procedure to generate Multiple lines of sight
density and velocity fields}
We briefly report  here the correlation procedure of Viel {\it et al.} (2002a).

Let us start by obtaining auto and cross-spectra for the 1D 
random fields which are defined along single or multiple LOS. 
In the Gaussian case these quantities fully determine the statistical 
properties of the generated fields.
If $\psi({\bf x})$ is a 3D random field with Fourier transform 
$\psi({\bf k})$ and 3D power spectrum $P(|k|)$, one can
define the LOS random field $\psi_\parallel(x_\parallel,{\bf x}_\perp)$
as the 1D Fourier transform  
\be
\psi_\parallel(k_\parallel| {\bf x}_\perp) \equiv \int {d^2 k_\perp
\over (2\pi)^2} e^{i{\bf k}_\perp \cdot {\bf x}_\perp}
\psi(k_\parallel,{\bf k}_\perp) \;. 
\ee
The cross-spectrum $\pi(|k_\parallel||r_\perp)$ for this LOS random field 
along parallel LOS, separated by a transverse distance $r_\perp$, is defined 
by
\be
\langle \psi_\parallel(k_\parallel|{\bf x}_\perp) 
\psi_\parallel(k^\prime_\parallel|{\bf x}_\perp + 
{\bf r}_\perp) \rangle = 2 \pi \delta_D(k_\parallel 
+ k^\prime_\parallel) \pi(|k_\parallel||r_\perp) \;, 
\ee
where $\delta_{\rm D}$ is the Dirac delta function and 
$\pi(k|r_\perp)$ can be related to the 3D power spectrum as follows 
\be
\pi(k|r_\perp) = \int {d^2 k_\perp \over (2\pi)^2} 
e^{i {\bf k}_\perp \cdot {\bf r}_\perp} P(\sqrt{k_\perp^2 + k^2}) \;. 
\label{eq:gencross}
\ee
Integrating over angles and shifting the integration variable yields
\be
\pi(k|r_\perp) = {1 \over 2 \pi} 
\int_k^\infty dq q J_0(r_\perp\sqrt{q^2 - k^2}) P(q) \;,
\label{eq:cross}
\ee 
where $J_0$ is the Bessel function of order $0$. 

In the limit of vanishing distance between the two LOS $J_0 \to 1$ and
the above formula reduces to the standard relation for the LOS (1D)
auto-spectrum in terms of the 3D power spectrum (Lumsden {\it et al.}
1989) 
\be 
p(k) \equiv \pi(k|r_\perp = 0) = {1 \over 2\pi}
\int_k^\infty dq q P(q) \;.
\label{eq:autosp}
\ee

Our next problem is how to generate the two random 
fields $\delta^{\rm IGM}$ and $v^{\rm IGM}$ in 1D 
Fourier space ($-\infty < k_\parallel < \infty$). These  
random fields have non-vanishing  cross-correlations but 
unlike their 3D Fourier space counterparts they cannot be related by simple 
algebraic transformations.  

We gan generally write any M-dimensional Gaussian random vector 
${\bf V}$ with correlation matrix ${\bf C}$ and components   
$c_{ij} = \langle V_i V_j \rangle$, 
as a linear combination of another  
M-dimensional Gaussian random vector ${\bf X}$ 
with diagonal  correlation matrix, which we can take as the identity 
${\bf I}$ without any loss of generality. 
The transformation involves the M $\times$ M  matrix ${\bf A}$, 
with components $\alpha_{ij}$, as follows:
${\bf V} = {\bf A} {\bf X}$. One  gets
${\bf C} = {\bf A} {\bf A}^T$, i.e. 
$c_{ij}  = \sum_k \alpha_{ik} \alpha_{jk}$. 
There is a slight complication because 
${\bf V}$ is a random vector ${\it field}$ 
defined in 1D Fourier space. We can, however,  extend 
the above formalism to  vector fields, assuming 
that ${\bf X}$ is a Gaussian vector
field with white-noise power spectrum,     
\be 
\langle X_i(k_\parallel)  X_j(k_\parallel^\prime) \rangle = 2 \pi 
~\delta_{ij}~\delta_D(k_\parallel + k_\parallel^\prime) \;
\ee
($\delta_{ij}$ is the Kronecker symbol). 
We then have $V_i = \sum_{j=1}^2 \alpha_{ij} X_j$ 
where 3 of the 4 $\alpha_{ij}$ components are determined by the 
conditions $\sum_{k=1}^2 \alpha_{ik} \alpha_{jk} = p_{ij}$.   
The remaining freedom (due to the symmetry of the original
correlation matrix) can be used to simplify the calculations. 
 
It is straightforward to extend our formalism to 
simulate the IGM properties along parallel LOS. 
Let  ${\bf V}(k_\parallel)$ and ${\bf W}(k_\parallel)$ be two 
1D Gaussian random vector fields obtained as in Section 2,
each with the same set of coefficients $\alpha_{ij}$  
but starting from two independent white-noise
vector fields ${\bf X}$ and ${\bf Y}$ 
(i.e. such that $\langle X_i Y_j \rangle =0$). 
Then both ${\bf V}$ and ${\bf W}$ have the
correct LOS auto-spectra by construction 
while their mutual cross-spectra vanish: 
$\langle V_i (k_\parallel) W_j(k^\prime_\parallel) \rangle
= 0$. 

Let us further define a new vector 
${\bf V}^\prime(k_\parallel|r_\perp)$ with components 
$V_i^\prime= \sum_{k=1}^2\left( \beta_{ik} V_k + \gamma_{ik} W_k
\right)$,
such that its auto and cross-spectra components are given by, 
\ba \langle V_i(k_\parallel) V_j(k_\parallel^\prime) \rangle & = &
\langle V^\prime_i(k_\parallel|r_\perp) V^\prime_j(k_\parallel^\prime|r_\perp)
\rangle = 2 \pi ~\delta_D (k_\parallel + k_\parallel^\prime)
p_{ij}(|k_\parallel|) \;, \nonumber \\
\langle V_i(k_\parallel) V^\prime_j(k_\parallel^\prime|r_\perp) 
\rangle & = &  
2 \pi ~\delta_D (k_\parallel + k_\parallel^\prime)
\pi_{ij}(|k_\parallel||r_\perp) \;. 
\ea

The vectors ${\bf V}$ and ${\bf V}^\prime$ will then represent
our physical IGM linear fields on parallel LOS at a distance 
$r_\perp$. They will be statistically indistinguishable from 
those obtained by drawing two parallel LOS separated 
by $r_\perp$ in a 3D realization of the linear IGM density
and velocity fields.   

The transformation coefficients are determined by the
equations $\sum_{k,\ell=1}^2 \left(\beta_{ik}\beta_{j\ell} + 
\gamma_{ik}\gamma_{j\ell} \right) p_{k\ell} = p_{ij}$ and
\linebreak$\sum_{k=1}^2 \beta_{ik} p_{kj} =  \pi_{ij}$. A complete set
of coefficients can be found in Viel {\it et al.} (2002a).

\end{document}